\documentclass[prl,twocolumn,showpacs,preprintnumbers,amsmath,amssymb,aps]{revtex4-1}

\usepackage{braket}
\usepackage{graphicx}
\usepackage{dcolumn}
\usepackage{bm}
\usepackage{setspace}
\usepackage{color}

\newcommand{\Yb}{\ensuremath{^{171}\mathrm{Yb}^+~}}
\newcommand{\Ba}{\ensuremath{^{138}\mathrm{Ba}^+~}}

\newcommand{\up}{\ensuremath{\left|\uparrow\right\rangle}}
\newcommand{\down}{\ensuremath{\left|\downarrow\right\rangle}}

\begin{document}

\title{Single-qubit quantum memory exceeding $10$-minute coherence time}

\author{Ye Wang$^{1}$, Mark Um$^1$, Junhua Zhang$^1$, Shuoming An$^1$, Ming Lyu$^1$, Jing -Ning Zhang$^1$, L.-M. Duan$^{1,2}$, Dahyun Yum$^{1*}$ \& Kihwan Kim$^{1*}$}

\affiliation{ $^1$ Center for Quantum Information, Institute for Interdisciplinary Information Sciences, Tsinghua University, Beijing 100084, P. R. China \\ 
$^2$ Department of Physics, University of Michigan, Ann Arbor, Michigan 48109, USA
}

\date{\today}

\begin{abstract}
A long-time quantum memory capable of storing and measuring quantum information at the single-qubit level is an essential ingredient for practical quantum computation and com-munication\cite{Ladd10,DuanMonroe12}. Recently, there have been remarkable progresses of increasing coherence time for ensemble-based quantum memories of trapped ions\cite{Bollinger91,Fisk95}, nuclear spins of ionized donors\cite{Saeedi13} or nuclear spins in a solid\cite{Zhong15}. Until now, however, the record of coherence time of a single qubit is on the order of a few tens of seconds demonstrated in trapped ion systems\cite{Langer05,Haffner05,Harty14}. The qubit coherence time in a trapped ion is mainly limited by the increasing magnetic field fluctuation and the decreasing state-detection efficiency associated with the motional heating of the ion without laser cooling\cite{Epstein07,Wesenberg07}. Here we report the coherence time of a single qubit over $10$ minutes in the hyperfine states of a \Yb ion sympathetically cooled by a \Ba ion in the same Paul trap, which eliminates the heating of the qubit ion even at room temperature. To reach such coherence time, we apply a few thousands of dynamical decoupling pulses to suppress the field fluctuation noise\cite{Khodjasteh13,Biercuk09,Zhong15,kotler2013nonlinear,Saeedi13,Souza11,haeberlen1976high}. A long-time quantum memory demonstrated in this experiment makes an important step for construction of the memory zone in scalable quantum computer architectures\cite{Kielpinski02,Hensinger15} or for ion-trap-based quantum networks\cite{DuanMonroe12,JKim13,Benjamin14}. With further improvement of the coherence time by techniques such as magnetic field shielding and increase of the number of qubits in the quantum memory, our demonstration also makes a basis for other applications including quantum money\cite{Wiesner83,Pastawski12}.
\end{abstract}


\maketitle

The trapped ion system constitutes one of the leading candidates for the realization of large-scale quantum computers\cite{Ladd10}. It also provides a competitive platform for the realization of quantum networks which combines long-distance quantum communication with local quantum computation\cite{DuanMonroe12}. One scalable architecture for ion-trap quantum computer is to divide the system into operation and memory zones and to connect them through ion shuttling\cite{Kielpinski02,Hensinger15}. For this architecture, the basic unit of operation zone has been demonstrated\cite{Home09, Hanneke10}. As the size of the system scales up, the needed storage time of the qubits in the memory zone will correspondingly increase. To keep the qubit error rates below a certain threshold for fault-tolerant computation, it is crucial to extend the coherence time of qubits. For the quantum network based on probabilistic ion-photon mapping\cite{duan2004scalable}, the basic units of ion-photon and ion-ion entanglement have been demonstrated\cite{blinov2004observation,Moehring07, Eschner14}. The required coherence time of qubits increases in this approach as the size of the system grows. A long-time quantum memory is therefore important for both quantum computation and communication\cite{DuanMonroe12,nielsen2010quantum}.

For trapped ion qubits, the main noise is not relaxation with time $T_1$ but instead dephasing with time $T_2^*$ induced by fluctuation of magnetic fields. The current records of single-qubit coherence time in trapped ion systems are around tens of seconds, achieved by using magnetic field insensitive qubits\cite{Langer05,Harty14} or decoherence-free-subspace qubits\cite{Haffner05}. The coherence time is mainly limited by motional excitations without cooling laser beam, as the latter would immediately destroys the quantum state of the qubit. Due to the motional heating, the wavepacket of the ion is expanding with time, which is subject to influence of nonuniform magnetic field. In addition, the motional heating reduces the count of fluorescence photons, which makes the qubit detection increasingly inefficient\cite{Epstein07,Wesenberg07}. Here, we completely eliminate the ion heating during long-time storage by the sympathetic cooling of a different species of atomic ions.

\begin{figure}[ht]
\centering
\includegraphics[width=0.5\textwidth]{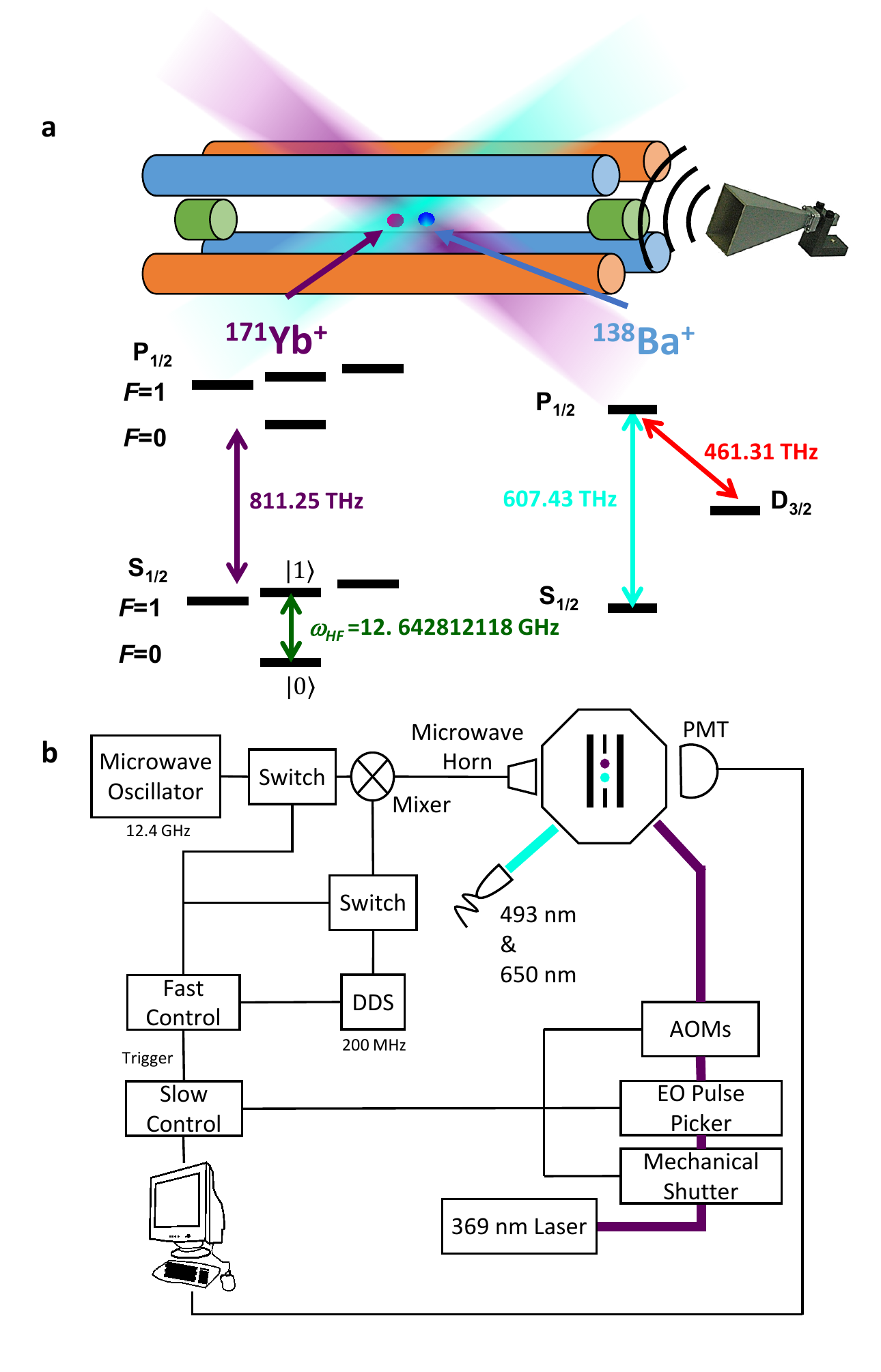}
\caption{\label{fig1}
\textbf{Experimental setup} \textbf{a}, Schematic diagram of a trapped-ion system with two species. We simultaneously trap $^{171}$Yb$^{+}$ and $^{138}$Ba$^{+}$ with distance of 10 $\mu$m in a linear Paul trap. The laser beams for $^{171}$Yb$^{+}$ and $^{138}$Ba$^{+}$ cover both ions so that initialization and detection of $^{171}$Yb$^{+}$ is not affected by the change of the ion position due to hopping in around every 5 minutes. We do not observe any difference in the strength of magnetic field and $\pi$ time of a single qubit gate for both positions. \textbf{b},
The microwave and laser control system. Microwave is generated by mixing a fixed 12.442812 GHz signal from an Agilent microwave oscillator and the signal around 200 MHz signal from a DDS with the capability of changing phase within 100 ns, which is controlled by FPGA. All sources are referenced to Rb clock. We apply a three-stage of switch systems for the 369 nm laser beam, which are AOMs, EO pulse picker and mechanical shutter (see Methods).}
\end{figure}

In the experiment, we use two species of ions, \Yb as the qubit ion and \Ba as the cooling ion, confined together in a standard Paul trap as shown in Fig. \ref{fig1}a. We choose \Ba ion as the refrigerator since it has similar mass to \Yb, which makes the sympathetic cooling efficient. The Doppler cooling laser and the repumping laser for \Ba are detuned by more than 200 THz from the corresponding transitions of \Yb ion, therefore the coherence of \Yb qubit is not affected by the cooling of \Ba ion (see Methods). The two hyperfine levels of \Yb ion in the $^{2}S_{1/2}$ manifold are used as a qubit represented by $\Ket{\downarrow} \equiv \Ket{F = 0 , m_{F} = 0} $ and $\Ket{\uparrow} \equiv \Ket{F = 1 , m_{F} = 0} $, which is separated by $12642812118+310.8 B^2$ Hz, where $B$ is magnetic field in Gauss. In experiment, we initialize the qubit state to $\Ket{\downarrow}$ by the standard optical pumping method and discriminate the state by the florescence detection scheme. We perform the coherent manipulation of qubit by applying the resonant microwave as shown in Fig. \ref{fig1}b.

With the help of sympathetic cooling, the remaining dominant factor of decoherence is the magnetic field fluctuation, which leads to phase randomization\cite{Biercuk09}. We can write the Hamiltonian of the qubit system as $H = \frac{\hbar}{2} (\omega_0+\beta(t)) \sigma_{\rm z}$, where $\omega_0$ is the splitting of the qubit, $\beta(t)$ is the random phase noise proportional to magnetic field fluctuation and $\sigma_{\rm z}$ is the Pauli operator. The accumulation of the random phase causes dephasing of the qubit state.

\begin{figure*}[ht]
\centering
\includegraphics[width=0.9\textwidth]{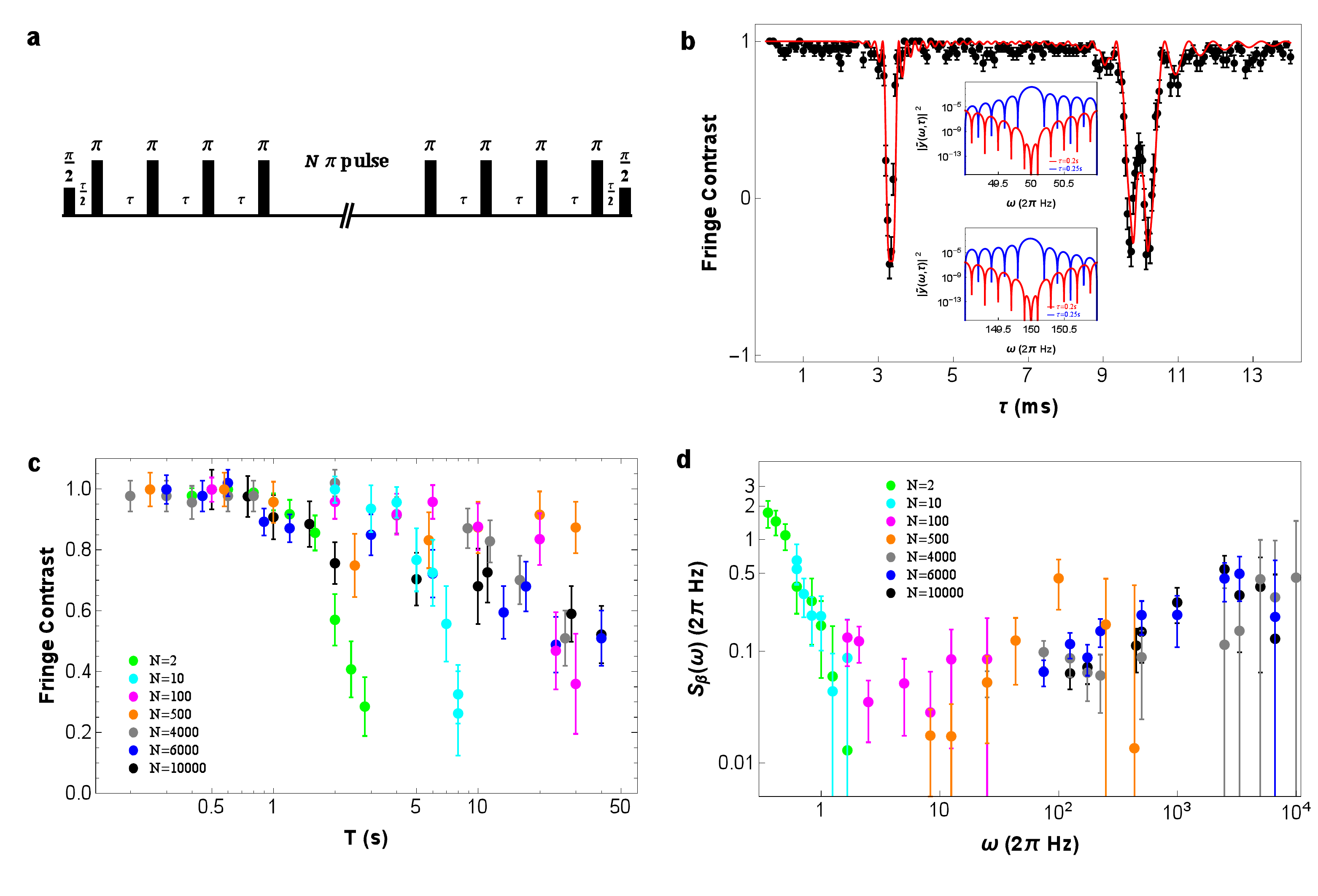}
\caption{\label{fig2}
\textbf{Measurement of noise spectrum of the system} \textbf{a}, Diagram of the CPMG sequence. All the $\pi$ pulses have the same phase of $\pi/2$. Interval between every $\pi$ pulses is $\tau$. \textbf{b}, The fringe contrast $\langle\cos(2F_N(T))\rangle$ as a function of the pulse interval for $31$ CPMG pulses. By fitting the data with Eq.~(\ref{DRC}), we obtain $B_{50 \rm Hz} = 18.3 \mu$ G, $B_{150 \rm Hz} = 57.5 \mu$ G. To suppress 50 Hz and 150 Hz noises with KDD$_{\rm xy}$, we choose $\tau$ = 200 ms instead of 250 ms, since the values of $|\tilde{y}|^2$ near 50 Hz and 150 Hz has around 10 orders of differences as shown in the inset, respectively. \textbf{c}, Ramsey contrasts $\langle\cos(2F_N(T))\rangle$ depending on the total evolution time for various numbers of pulses as $N =$ 2, 10, 100, 500, 4000, 6000 and 10000. The data are normalized to the contrast of continuous application of the same number of pulses. Error bars are the standard deviations of 50 to 200 repetitions. \textbf{d}, Analyzed noise spectra from the coherence decay shown in \textbf{c} by using Eq.~(\ref{chi}). The smallest noise level is located at the range of ($2\pi$) $2 \sim 100$ Hz.}
\end{figure*}

A standard technique to preserve the qubit coherence against random phase noise is dynamical decoupling, which expands Hahn spin echo into a multi-pulse sequence\cite{Khodjasteh13,Biercuk09,Zhong15,kotler2013nonlinear,Saeedi13,Souza11,haeberlen1976high}. Since the performance of the sequence is sensitive to the characteristics of noise environment, we study the noise spectrum of our system, which guides us to choose the proper dynamical decoupling pulses. In the experiment, we probe the specific frequency part of the noise spectrum by monitoring the response of the qubit under the specific sequence\cite{bylander2011noise}. The initial state of qubit $\ket{\psi(0)}$ under a dynamical decoupling sequence in noisy environment evolves according to $\ket{\psi(T)} = e^{i F_N (T) \sigma_{\rm z}} \ket{\psi(0)}$, where $F_{N} (T) = \sum_{i=0}^{N} (-1)^{i+1} \int_{\tau_i}^{\tau_{i+1}}dt \beta(t)$, $T$ is the total evolution time, $\tau_{i}$ is the time stamp of the $i$-th $\pi$ pulse. We obtain the information of $F_N(T)$ by measuring the contrast of Ramsey fringe as $\langle\cos(2F_N(T))\rangle$, which we define as the coherence of the qubit (see Methods). We set magnetic field strength to 8.8 G and apply CPMG (Carr, Purcell, Meiboom and Gill)\cite{haeberlen1976high} sequence shown in Fig. \ref{fig2}a to measure the noise spectrum of the environment.

In our system, the dominant components of noise are at $50$ Hz and its harmonics coming from the power line, which is modeled as the sum of discrete noises $\tilde{\beta}(\omega) = \sum_{k=1}^{d} \beta_k \delta(\omega-\omega_k)$. Here $\beta_k$ is the strength of the noise. The Ramsey contrast of the final state becomes\cite{kotler2013nonlinear}
\begin{eqnarray}
\langle\cos(2F_N(T))\rangle = \prod_{k=1}^d J_0(|\beta_k \tilde{y}(\omega_k,T)|),
\label{DRC}
\end{eqnarray}
where $J_0$ is the 0th of Bessel function, $\tilde{y}(\omega,T) = \frac{1}{\omega} \sum_{j=0}^{N} (-1)^j \left[e^{i \omega \tau_j}- e^{i \omega \tau_{j+1} }\right]$, $\tau_0 = 0$, $\tau_{N+1} = T$. With $31$ pulses, the resultant coherence as a function of the pulse interval is shown in Fig. \ref{fig2}b. By fitting, we obtain the discrete noise spectrum as $B_{50\rm Hz} = 18.3$ $\mu$G, $B_{150\rm Hz} = 57.5$ $\rm \mu$G with no significant components at other frequencies.

We further study the rest of the noise spectrum by using the continuous noise model.
Given an arbitrary noise spectrum $S_{\beta}(\omega)$, the Ramsey contrast is given by
$e^{-\chi(T)}$, where $\chi(T)$ is written as\cite{Biercuk09}
\begin{equation}
\chi(T) = \frac{2}{\pi}\int_0^\infty S_\beta (\omega) |\tilde{y}(\omega,T)|^2 d \omega.
\label{chi}
\end{equation}
Here the function $|\tilde{y}(\omega,T)|^2$ can be viewed as a bandpass filter with the center frequency of $\frac{1}{ 2\tau}$ and the width proportional to $\frac{1}{2\pi T}$, where $\tau$ is the pulses interval shown in Fig. \ref{fig2}a. As shown in Fig. \ref{fig2}c, we observe the Ramsey contrasts depending on the total evolution time $T$ with various total number of pulses as $N =$ 2, 10, 100, 500, 400, 6000 and 10000. Applying Eq.~(\ref{chi}) to the results of Fig. \ref{fig2}c, we obtain $S_\beta(\omega)$ as shown in Fig. \ref{fig2}d. We observe sharp increase of noise strength below (2$\pi$) 2 Hz, which is consistent with the result of flux gauge measurement. We also observe slow increase of the noise above (2$\pi$) 100 Hz, which could come from our ground line for current sources generating the magnetic field.

In order to extend coherence time with the CPMG type of sequence under our current noise environment, we carefully choose the interval of the pulse sequence. We can locate the bandpass frequency between 2 Hz and 100 Hz, where the noise spectrum has lowest value, which means $5< \tau < 250 $ ms. In this range, a larger $\tau$ is preferred, which reduces the number of pulses and leads to smaller accumulation of gate errors. We choose the pulse interval $\tau$ properly so that the noise components at $50$ Hz and $150$ Hz are suppressed to a negligible level. Considering all the factors, the optimal $\tau=200$ ms.

\begin{figure}[ht]
\centering
\includegraphics[width=0.43\textwidth]{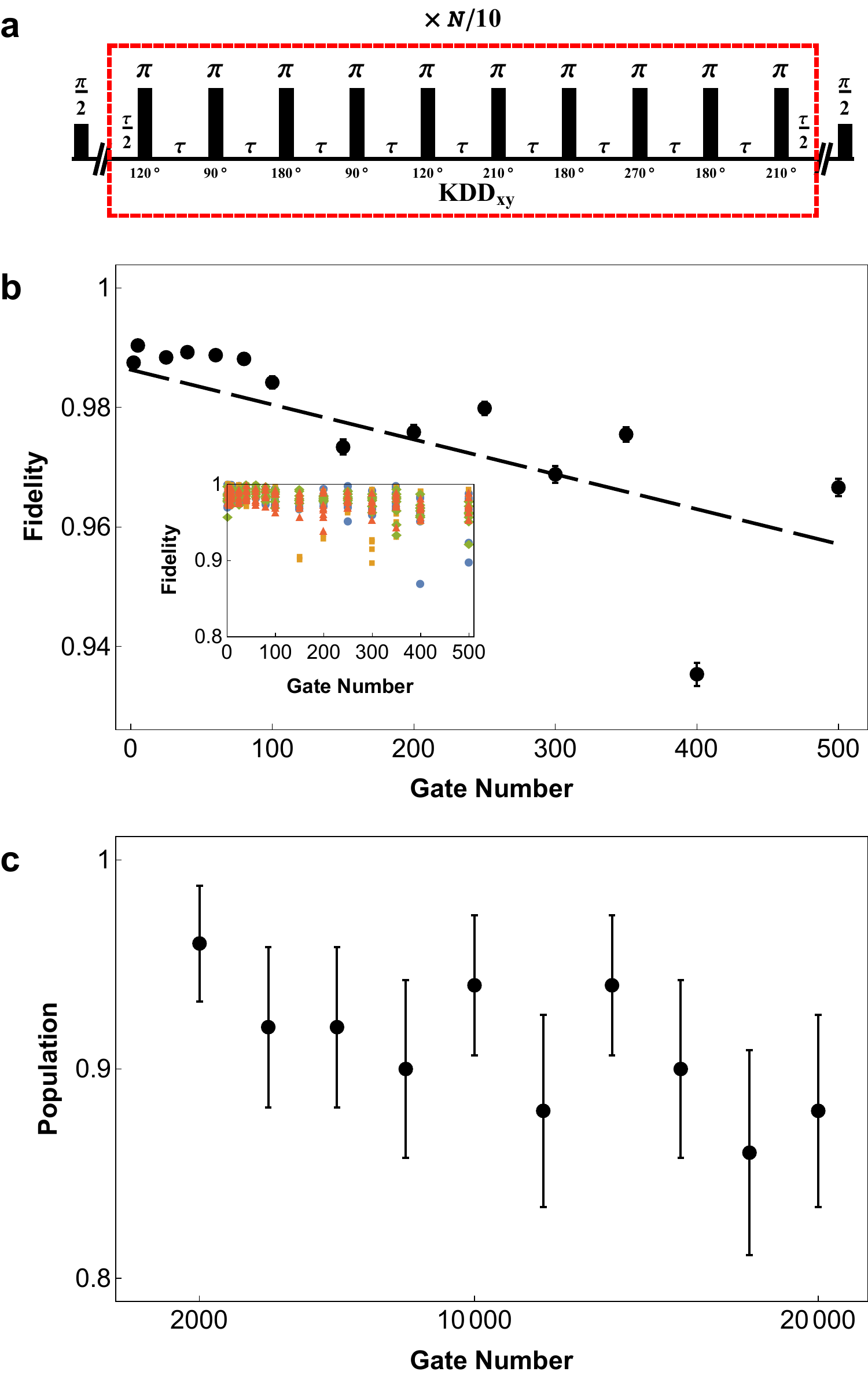}
\caption{\label{fig3}{}
\textbf{KDD$_{\rm xy}$ sequence and gate fidelity} \textbf{a}, Diagram of KDD$_{\rm xy}$ sequence. Knill pulse consists of five $\pi$ pulses with angles $\phi+\pi/6,\phi,\phi+\pi/2,\phi,\phi+\pi/6$. The KDD$_{\rm xy}$ consists of two Knill pulses equally spaced with $\phi = \pi/2$ and $\phi = \pi$. We apply $N/10$ sets of the KDD$_{\rm xy}$, which leads to the total number of $N$ pulses. Note that $N$ should be a multiple of 20 to form an identity operation\cite{Souza11}. \textbf{b}, Randomized benchmarking test of the single qubit gates. We use 32 settings of random sequences followed by Ref.\cite{knill2008randomized}. At each setting we repeat 500 times. The inset shows the results of 32 different settings. The black dots in \textbf{b} represent the average values of 32 settings with the error bar of standard deviation. By fitting, we obtain the gate fidelity as $99.994\pm0.002\%$.  \textbf{c}, We continuously apply the KDD$_{\rm xy}$ pulses on the initial \up~state to test its performance. We observe more than $85\%$ population back to the initial \up~state after the application of 20000 $\pi$ pulses.}
\end{figure}

In experiment, we use the KDD$_{\rm xy}$ \cite{Souza11,Zhong15} sequence to store arbitrary quantum states of single qubit. KDD$_{\rm xy}$ is a robust dynamical decoupling sequence insensitive to imperfections of gate operations including flip-angle errors and off-resonance errors. As shown in Fig. \ref{fig3}a, KDD$_{\rm xy}$ can be considered as an extension of the CPMG sequence (see Methods). Our single qubit gate fidelity is measured to be $99.994\pm0.002\%$ by randomized benchmarking method\cite{knill2008randomized} as shown in Fig. \ref{fig3}b. We test the robustness of KDD$_{\rm xy}$ by continuously applying it to \up~state. We have more than $85\%$ population back to the initial \up~state after the application of $20000$ pulses as shown in Fig. \ref{fig3}c.

\begin{figure}[ht]
\centering
\includegraphics[width=0.45\textwidth]{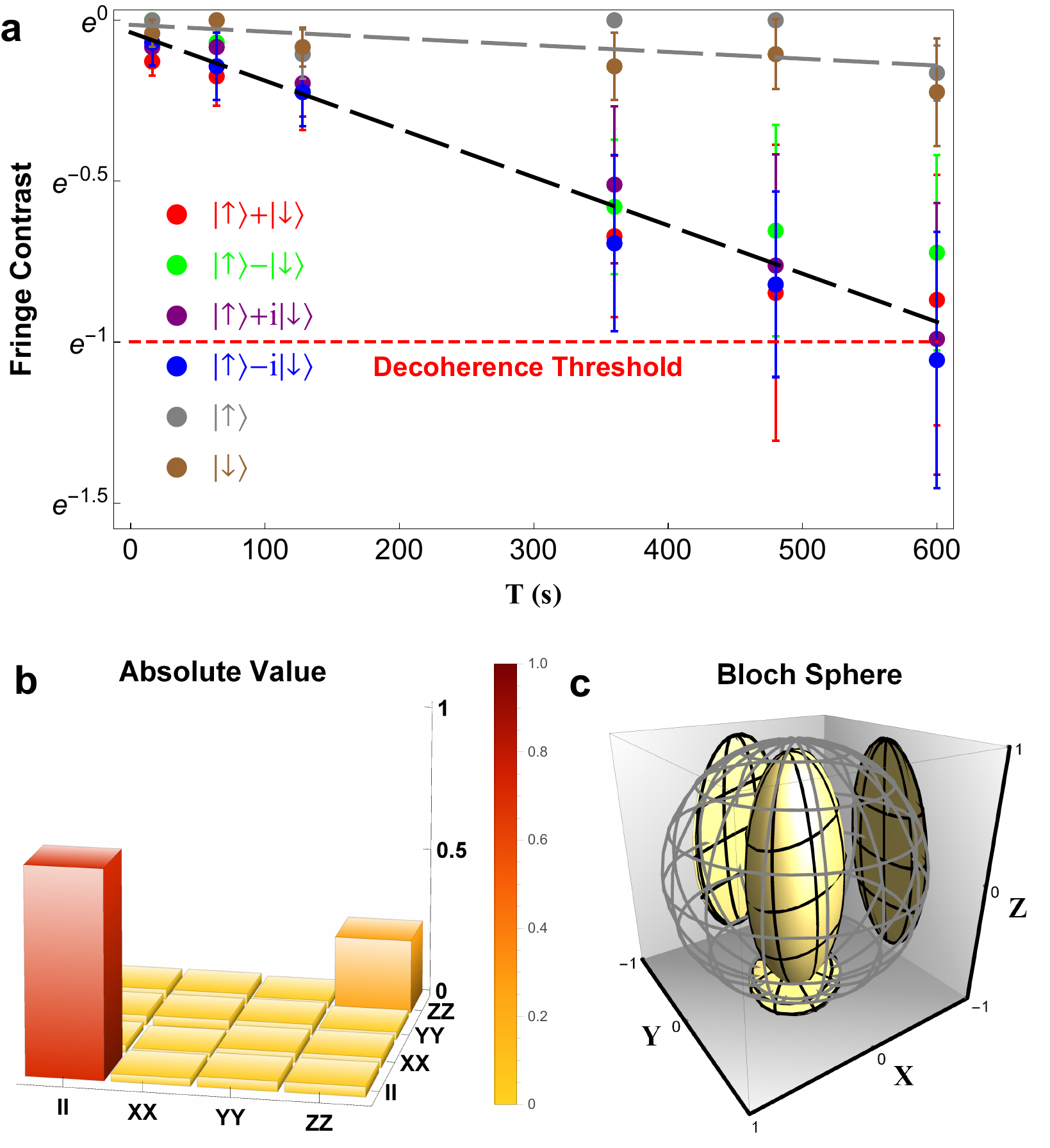}
\caption{\label{fig4}
\textbf{Coherence time measurement and quantum process tomography} \textbf{a}, The coherence time of six different initial states. For the state \up~and \down, the coherence time is $4744.72\pm1762.33$ s. For the other four initial states,  the coherence time is $666.9\pm16.6$ s. \textbf{b}, The result of quantum process tomography for the duration of 8 minutes. Here the tomography is obtained from the entire process including initialization, storage and detection. Identity is the dominant diagonal part with $\chi_{\rm II} = 0.699\pm 0.058$. \textbf{c}, The transformation from the initial states lying on the meshed surface to the final states lying on the solid surface after 8 minutes of storage.}
\end{figure}

For the measurement of single-qubit coherence time, we set the magnetic field to $3.5$ G. We measure the coherence time of single-qubit memory with 6 different initial states as $\Ket{\uparrow}$, $\Ket{\downarrow}$, $1/\sqrt{2}(\Ket{\uparrow}+\Ket{\downarrow})$, $1/\sqrt{2}(\Ket{\uparrow}+i\Ket{\downarrow})$, $1/\sqrt{2}(\Ket{\uparrow}-\Ket{\downarrow})$, $1/\sqrt{2}(\Ket{\uparrow}-i\Ket{\downarrow})$. In Fig. \ref{fig4}a, we see no significant relaxation for the initial states of $\Ket{\uparrow}$ and $\Ket{\downarrow}$ under KDD$_{\rm xy}$ sequence. The reduction of their contrasts mainly comes from the accumulation of gate errors. The other four initial states have a similar decoherence rate, which corresponds to the coherence time of $666.9\pm16.6$ s from the exponential decay fitting. The demonstrated coherence time is primarily limited by non-vanishing noise strength at the bandpass filter frequency of $2.5$ Hz. We also perform the process tomography on our dynamical decoupling process for the duration of 8 minutes. The results shown in Fig. \ref{fig4}b and \ref{fig4}c demonstrate that the process is approximated as an identity operator with the main error corresponding to pure dephasing, in agreement with our knowledge about the hyperfine qubit of the trapped ion system.

We note that our measured coherence time over $600$ s is not fundamentally limited. The coherence time can be further increased by decreasing $S_\beta(\omega)$ at the bandpass frequency, which can be achieved by installing a magnetic field shield or by using a magnetic field insensitive qubit at zero crossing regime. The number of memory qubits can be increased with sympathetic cooling for quantum cryptographic applications including quantum money\cite{Wiesner83,Pastawski12}.

\subsection{Acknowledgment}
This work was supported by the National Key Research and Development Program of China under Grants No. 2016YFA0301900 (No. 2016YFA0301901), the National Natural Science Foundation of China 11374178, 11504197 and 11574002.

\section{Method}
\subsection{Control System}

We use two FPGA boards to separately control the laser system and the microwave system. Two parts are synchronized in less than 10 ps of time jitter. The leakage of 369 nm laser beams greatly limits the coherence time of the qubit of \Yb ion. To block the 369 nm laser beams for the Doppler cooling, initializing and detecting the quantum state of \Yb ion, we use three stages of switches: AOMs (Acousto-Optic Modulator), EO (Electro-Optic) pulse picker and mechanical shutter. The AOM switches leave a few nW of laser beam leakage, the EO pulse picker provides 200 times of attenuation for the leakage and the mechanical shutter blocks the laser beam completely with about 50 ms of response time. With only the AOM switches, we observe a maximum of 50 ms coherence time. The coherence time is increased to 10 s by applying EO pulse picker and dynamical decoupling sequence. Therefore it is necessary to use the mechanical shutter to completely block the laser beam. The resonant microwave with the frequency of 12.642812 GHz is generated by mixing the signal of a microwave oscillator and the signal of around 200 MHz of a DDS (Direct Digital Synthesizer). The DDS is used to change the phase of the microwave within 100 ns, which is controlled by the FPGA. All sources are referenced to a Rb clock.

\subsection{Limitation of coherence time due to the cooling laser beams for $^{\bf 138}$Ba$^{\bf +}$}
We calculate the scattering rate between $^{2}S_{1/2}$ manifold and $^{2}P_{1/2}$, $^{2}P_{3/2}$ manifold of \Yb due to the cooling laser beams, 493 nm and 650 nm, for \Ba ion. The scattering rate is calculated by the model of Raman scattering\cite{ozeri2005hyperfine,uys2010decoherence,campbell2010ultrafast}, which is given as 
\begin{eqnarray}
\Gamma_{\rm scat} &=& \frac{g^2 \Gamma}{6} \left( \frac{1}{\Delta_{\rm D1}^2}+ \frac{2}{(\Delta_{\rm D1}+\Delta_{\rm fs})^2}\right), 
\end{eqnarray}
where $\Gamma \approx 2\pi \times 20$ MHz, $g = \frac{\Gamma}{2} \sqrt{ \frac{I}{2I_{\rm sat}}}$, $\Delta_{\rm fs}\approx 2\pi \times 100$ THz. For the 493 nm laser, the parameters are as follows: laser power $P = 40$ $\mu$w, beam waist $\omega = 22$ $\mu$m,  $I_{493} = 49.6 I_{\rm sat}$ and $\Delta_{\rm D1}$= 203.8 THz, which yields a scattering rate of $3.8 \times 10^{-8}$ Hz. For the 650 nm laser, the parameters are as follows: laser power $P = 10$ $\mu$w, beam waist $\omega = 20.7$ $\mu$m,  $I_{650} = 14.0 I_{\rm sat}$ and $\Delta_{\rm D1}$= 349.9 THz, which yields a scattering rate of $1.5 \times 10^{-8}$ Hz. Therefore, 493 nm and 650 nm laser beams have no effect on the coherence of \Yb ion hyperfine qubit during hours of storage.

\subsection{Magnetic field fluctuation} The second order Zeeman effect of the \Yb qubit transition\cite{fisk1997accurate} is described by $\Delta f_{2\rm OZ}=K \langle B^2\rangle$ Hz, $K = 310.8$ Hz/G$^2$. In the Hamiltonian $H = \frac{\hbar}{2}(\omega_0+\beta(t))\sigma_{\rm z}$, where $\omega_0$ and $\beta(t)$ are given by 
\begin{eqnarray}
\label{noiseModel1}
 &&\omega_0 = \omega_{\rm{HF}}+K(B_{\rm x}^2+B_{\rm y}^2+B_{\rm z}^2),\\
\label{noiseModel2}
 &&\beta(t) = K(2 B_{\rm x} \langle b_{\rm x}(t)\rangle +2 B_{\rm y} \langle b_{\rm y}(t)\rangle + 2 B_{\rm z} \langle b_{\rm z}(t)\rangle \nonumber\\
 &&+\langle b_{\rm x}(t)^2\rangle +\langle b_{\rm y}(t)^2\rangle +\langle b_{\rm z}(t)^2\rangle ),
\end{eqnarray}
and $\omega_{\rm{HF}}$ is the hyperfine splitting, $B_{\rm x}$, $B_{\rm y}$, $B_{\rm z}$ represent the average values of the magnetic field in three directions and $b_{\rm x}(t)$, $b_{\rm y}(t)$, $b_{\rm z}(t)$ describe fluctuations in the corresponding directions. The fluctuations are believed to come from the environment and the power sources generating the magnetic field. Since $b_{\rm x}(t)$ is much smaller than $b_{\rm y}(t)$ and $b_{\rm z}(t)$ according to the flux gauge measurement, we put the magnetic field along x-axis. In the experiment, we set $B_{\rm z} = 0$, $B_{\rm y} = 0$ and $B_{\rm x} = 3.5$ G, which is the smallest magnetic field strength to maintain detection efficiency in our system. To study noise spectrum of our system, we use the setting of $B_{\rm z} = 0$, $B_{\rm y} = 0$ and $B_{\rm x} = 8.8$ G.

\subsection{Dynamical decoupling}
In the experiment, CPMG is used for the study of noise environment and KDD$_{\rm xy}$ is used for storage. Here we show the filter function of KDD$_{\rm xy}$ is the same as that of CPMG. Since all the rotation axes we use are in xy-plane, we write the rotation as
\begin{eqnarray}
D_{\phi}(\gamma) &=& D_{\rm z}(\phi)D_{\rm x}(\gamma)D_{\rm z}(-\phi)\nonumber \\
&=& e^{-\frac{i}{2}\sigma_{\rm z}\phi} e^{-\frac{i}{2}\sigma_{\rm x}\gamma} e^{\frac{i}{2}\sigma_{\rm z}\phi}\label{expand},
\end{eqnarray}
where $\phi$ is the angle between the rotation axis and $x$ axis and $\gamma$ is the rotation angle along the axis. In dynamical decoupling sequences, $\gamma$ is always $\pi$, therefore, $D_{\rm x}(\pi) =  \cos(\pi/2)+i\sigma_{\rm x} \sin(\pi/2) = i\sigma_{\rm x}$. Below we will only use $\sigma_{\rm x}$ because the factor $i$ is an irrelevant global phase.

We start from initial state $\ket{\psi(0)} = (\ket{\downarrow}+i \ket{\uparrow})/\sqrt{2}$, the final state is shown as
\begin{eqnarray}
&&\ket{\psi(T)} =\tilde{R}(T)\ket{\psi(0)}, \\
&&\tilde{R}(T) = e^{-i\sigma_{\rm z} \int_{\tau_{N}}^{\tau_{N+1}}\beta(t) dt} D_{\phi_N}(\pi) \cdots \nonumber\\
&&D_{\phi_2}(\pi)e^{-i\sigma_{\rm z} \int_{\tau_1}^{\tau_2}\beta(t) dt} D_{\phi_1}(\pi)e^{-i\sigma_{\rm z} \int_{\tau_0}^{\tau_1}\beta(t) dt}\label{R1},
\end{eqnarray}
where $\tilde{R}(T)$ is the evolution during total time $T$, $N$ is the total number of $\pi$ pulses, $\tau_i$ is the time stamp of the $i$th $\pi$ pulse, $\tau_0 = 0$, $\tau_{n+1} = T$ and $\phi_i$ is the phase of the $i$th $\pi$ pulse. Since $N$ is even for our dynamical decoupling sequence and only $\sigma_{\rm x}$ operations in $\tilde{R}(T)$ flip the spin, we can obtain
\begin{eqnarray}
\tilde{R}(T)  &=& e^{iF_N(T)\sigma_{\rm z}} \label{R2},
\end{eqnarray}
where $F_N(T)$ is a phase term which is obtained by expanding $D_{\phi}(\pi)$ in Eq.~(\ref{R1}) with Eq.~(\ref{expand}) as
\begin{eqnarray}
F_N(T) = \sum_{i=0}^N{(-1)^{i+1}\int_{\tau_{i} }^{\tau_{i+1} }\beta(t)}+\sum_{i=1}^N{(-1)^{i+1}\phi_i}.
\end{eqnarray}
According to Eq.~(\ref{R2}), the Ramsey fringe contrast we measure is\cite{uhrig2008exact}
\begin{eqnarray}
\bra{\psi(T)}\sigma_{\rm y}\ket{\psi(T)} &&= \bra{\psi(0)}\tilde{R}(T)^\dag \sigma_{\rm y}  \tilde{R}(T)  \ket{\psi(0)} \nonumber\\
&&= \langle \cos(2F_N(T))\rangle.
\end{eqnarray}

For normal dynamical decoupling sequences, like CPMG, XY-16 and KDD$_{\rm xy}$, we have $\sum_{i=1}^N{(-1)^{i+1}\phi_i} = k \pi$, where $k$ is an integer. Especially for KDD$_{\rm xy}$, $N$ is a multiple of 20 and $\phi_i$ is cycling in the order of $\pi/6$, $0$, $\pi/2$, $0$, $\pi/6$, $2\pi/3$, $\pi/2$, $\pi$, $\pi/2$, $2\pi/3$. So $\sum_{i=1}^{20} {(-1)^{i+1}\phi_i} = -\pi$, which means $\langle \cos(2F_N(T))\rangle$ is not affected by $\sum_{i=1}^N{(-1)^{i+1}\phi_i}$. So KDD$_{\rm xy}$ and CPMG with same interval and pulse number have the same evolution. We can simplify $\langle \cos(2F_N(T))\rangle$ with discrete and continuous model according to Ref.\cite{kotler2013nonlinear,Biercuk09}. 

\bibliographystyle{apsrev4-1}
%


\end{document}